# Tremor Reduction for Accessible Ray Based Interaction in VR Applications

Dr Corrie Green[1]

Robert Gordon University

c.green1@rgu.ac.uk

Dr Yang, Jiang

Robert Gordon University

y.jiang2@rgu.ac.uk

Dr John, Isaacs

Dean, Robert Gordon University

j.p.isaacs@rgu.ac.uk

Dr Michael, Heron

Chalmers University of Technology

heronm@chalmers.se

August 10th, 2022

## 1 ABSTRACT

Comparative to conventional 2D interaction methods, virtual reality (VR) demonstrates an opportunity for unique interface and interaction design decisions. Currently, this poses a challenge when developing an accessible VR experience as existing interaction techniques may not be usable by all users. It was discovered that many traditional 2D interface interaction methods have been directly converted to work in a VR space with little alteration to the input mechanism, such as the use of a laser pointer designed to that of a traditional cursor. It is recognized that distance-independent millimetres can support designers in developing interfaces that scale in virtual worlds. Relevantly, Fitts law states that as distance increases, user movements are increasingly slower and performed less accurately. In this paper we propose the use of a low pass filter, to normalize user input noise, alleviating fine motor requirements during ray-based interaction. A development study was conducted to understand the feasibility of implementing such a filter and explore its effects on end users' experience. It demonstrates how an algorithm can provide an opportunity for a more accurate and consequently less frustrating experience by filtering and reducing involuntary hand tremors. Further discussion on existing VR design philosophies is also conducted, analysing evidence that supports multisensory feedback and psychological models. The completed study can be downloaded from GitHub [1].

*Keywords* Virtual Reality · Ergonomics · Accessibility · User Interface · User Experience

## 2 INTERFACE ERGONOMICS

Areas of interaction that are not within the arm's reach of a user and will require physical movement to the location of the anchored interface. During prolonged periods this can not only result in fatigue but also prevent those with mobility restrictions from reaching the interface to interact. It is therefore important to consider constraints during development to support users' ergonomic ability.

Considering user depth perception and headset field of view, locating the optimal spatial area for VR interaction with a motion controller can start to be conceptualized. The below section further explores work completed in this space with the development of a graphic to support VR researchers in future VR interface decision-making. Whereby overlaying all the figures, we can establish a baseline for where content should be rendered in VR, for a comfortable experience. Such an interface developed from the solution provided includes the Modular 3D Interface for Multimodal input [2]. Here a 3D interface shown in Figure 2 allows users to interact with distance being close to the users body for interaction.

---

[1] https://orcid.org/0000-0003-0404-3668
https://github.com/corriedotdev/vr-tremor-reduction

## 2.1 Accessible Rotation

For understanding how head rotation can influence user comfort, Samsung conducted studies[3] that measured the range of comfort levels from users while turning their heads to the side whilst wearing a VR headset.

The formula below is using figures found in the study. It shows the subjective perception of comfort from the participant's head rotation with 30 degrees being a comfortable rotation range and 55 being the maximum before neck discomfort. However, every HMD has a different horizontal and vertical field of view (FOV) therefore it should be considered an estimate of user comfort range during head rotation [4].

$$Min\ Comfort\ Angle = \left(\frac{FoV}{2}\right) + 30$$

(1)

$$Max\ Comfort\ Angle = \left(\frac{FoV}{2}\right) + 55$$

(2)

Formula to calculate comfort ranges from a study on subjective comfort head ranges.

## 2.2 Accessible Reach

Studies into stereoscopic perception have shown that 20 meters is the maximum distance humans perceive depth [5]. This also coincides with a study [3] where users placed in a VR HMD could see strong 3D from 1m to 10m ranges. With depth perception falling off between 10m to 20m, ultimately resulting in a 2D visual with no depth. Using this data, you can start to understand where 3D-rich information can be placed for a user and develop areas of the virtual space for specific user engagement.

The immediate area surrounding the user should never contain any interactable content as most of the space is taken up by the user's physical torso. At 0.5m users would likely feel discomfort due to the high potential for being cross-eyed so is classified as personal space. When taking into consideration that VR headsets also are relatively large on the head, over prolonged periods there can be further neck discomfort when looking directly down or up, therefore personal space should be left with no interaction.

However, HMD hardware focal distances can also provide insight into existing interaction areas. The Meta Quest 2 HMD has a fixed focal distance of 1.3m. Known as the vergence-accommodation conflict, users' eyes must therefore try accommodating for the fixed focal distance by converging or diverging to a point in screen space. This design decision by Meta further supports the placement of objects at around the 1m range due to being in the optimal focal range for the user's hand movements. Until VR displays support varifocal lenses, we should continue to consider the clarity of objects in this range.

It is then suggested that the main hand interaction is between 0.5m to 1m from the user's center, where 3D depth perception is at its clearest, without cross-eye discomfort, and with existing HMD lenses supporting clear focus. Furthermore, a user's arm span is on average the same as a user's height. It would therefore be possible to scale this template algorithmically down or up accordingly based on user height while wearing the headset, however, there would need to be an additional study that takes comfortable user arm reach into consideration for a more accurate result. Arguably this zone is optimal for interactable components due to the comfortable legibility of presented text, with an evaluation of cockpit display interactions discussing values where it was said that the minimum comfortable viewing is at 35-50cm [6]. With supporting values in office ergonomics [7] where monitors are about 70cm away from the user's eye gaze.

Tremor Reduction for Accessible Ray Based Interactions in VR Applications

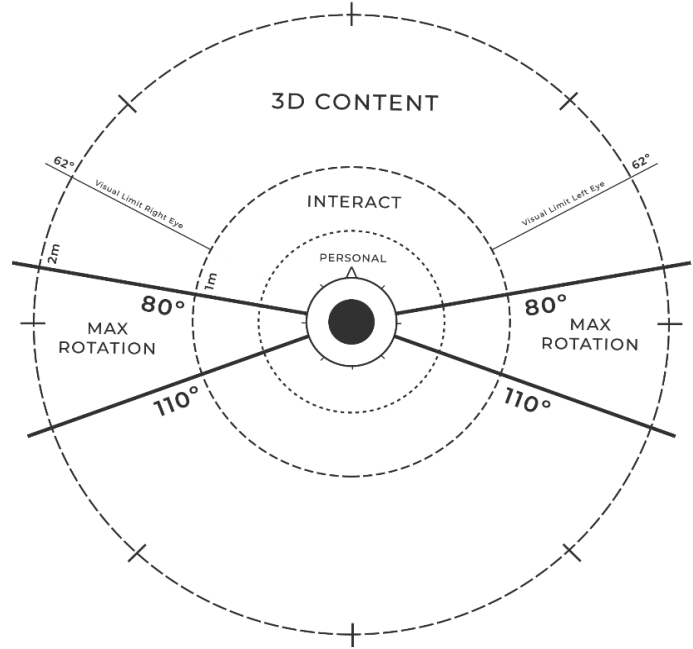

*Figure 1 Floor graphic showing interaction areas for VR experiences, a new contribution from this paper*

The main 3D content zone is therefore between 1m and 2m as depth perception is still perceived and current fixed focal distance headset displays support this area. Objects placed in this area are too far away from the user to interact without moving in virtual or physical space.

The Graphic shown above in Figure 1 takes the methods described in this section and combines them into a single visualization that can be placed on the floor surface of the virtual environment being developed. Outlining the areas that developers should use for guidance when placing static objects in the scene, or dynamic objects where the user's gaze will be predetermined based on range. Figure 2 demonstrates a developed 3D GUI system using this graphic as a key part of its accessible first design.

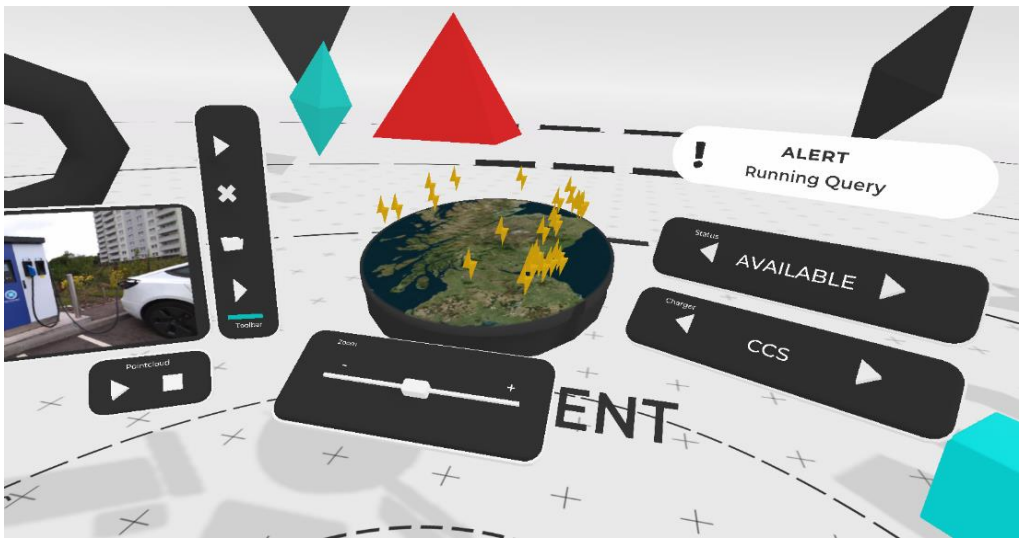

*Figure 2 A developed modular 3D VR GUI using the outlined constraints demonstrating spatial and diegetic interface elements*

## 2.3 Evaluation Procedure

A fully supervised in-person study was conducted with 36 participants aged 18-55 who were recruited from the public, faculty, and students. These participants identified having no movement disabilities and as far as they were aware free from neuromuscular disorders. Data from 6 participants were disregarded due to being unable to complete the tasks due to being unable to wear their glasses with the headset on and noted they have eye stigmatisms (N=4) with the remainder of participants (N=2) being unable to complete the tasks due to unable to see the interface elements clearly at various ranges.

Two UI elements were selected to be tested with 30 users, using a laser pointer input from the Unity XR framework. A 2D numpad with values 0-9 and a 2D slider with input from 0-100 were implemented into a Unity scene created from the previous simulation discussed as the filters proof of concept was already validated.

As tremors increase during fine motor requirements and with Fitts law being taken into consideration discussed in Section 2.6, users would be presented with these generated UI elements for interaction at increased ranges of 1m, 5m, and 10m. Users are tasked with inputting 4 randomly generated integers into the numpad at each range in their own time after pressing the start button. Once they have correctly entered the value, they will return to the now stop button to stop the timer. The same process is carried out for the slider; however, the generated value was between 20-100 to prevent users from clustering input at lower values of the slider, where its handle is initialized at zero. If the number 5 was generated the slider anchor only needs to be moved 5cm. The slider is 1m in length – left to right width - and represents 1cm per input value throughout the tests at different ranges, the numpad sizing is also the same throughout to focus control measures on the efficacy of the filter.

At first, the participant was introduced to the environment after adjusting the headset to a comfortable position and appropriate IPD value. Mistakes made by the user during integer selection using the numpad were rectifiable by the user using a backspace button. Users were shown the slider and numpad at 1m initially and were able to practice input once in their own time before the study started.

To control task order effects, a counterbalance is used in which participants have the filter on or off at the start of their task, randomly assigned 50% of the time. Only when each task at the 3 ranges was completed with the numpad and slider, they would complete the same tasks again but with the filters toggle value inverted.

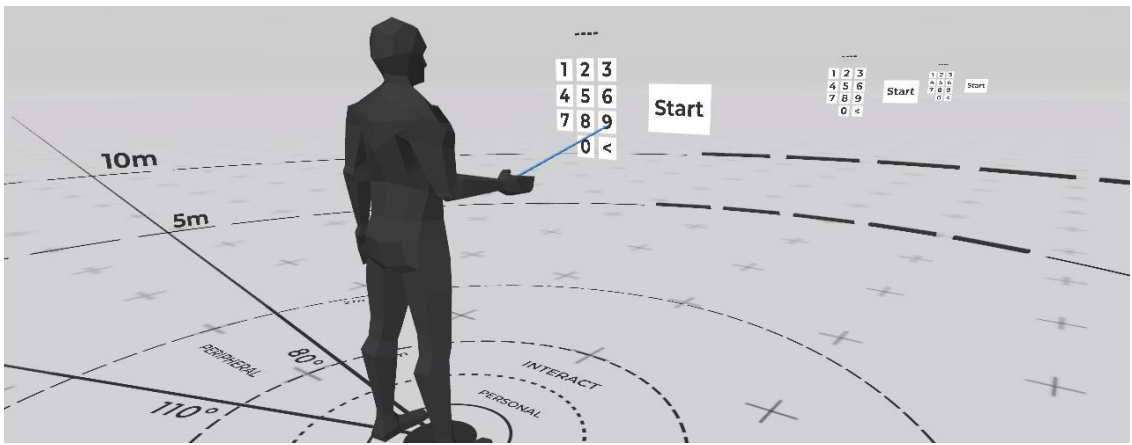

*Figure* **3** *Users test environment with the numpad and slider display at various ranges*

### 2.3.1 Results

In the following study, we investigated the null hypothesis that task completion times show no improvement with the 1-Euro filter enabled. A paired t-test was conducted to assess the null hypothesis that there is no significant difference in task completion times with the filter enabled, only where the p-value is less than the significance level which was set at α = 0.05.

*2.3.2 Slider*

Results indicate that at 10m ranged slider tasks there was a significant improvement in task completion time with the filter enabled t(29)= 3.1750, p=0.0035. Therefore, we reject the null hypothesis that task completion times show no improvement with the filter enabled, and find task completion time was on average 10.57 seconds faster with the filter enabled at 10m ranged tasks. This correlates to an average of 42.83% improvement in task completion time for users during 10m slider tasks, as shown in Table 2.

However we found minimal difference in completion time for 5m ranged tasks showing t(29)= 1.8518, p=0.0742 where we therefore accept the null hypothesis. This resulted in an average completion time of only 4.11 seconds faster with the filter enabled.

Similarly at 1m ranged tasks, we find no significance in completion time t(29)= 0.7833, p=0.4397 where we again accept the null hypothesis with the average task completion time only 0.55 seconds faster.

|  | 1m | 5m | 10m |
|---|---|---|---|
| **p-value** | 0.4397 | 0.0742 | 0.0035 |

*Table* **1** *Slider task completion p-value at given ranges, with a significance level at 0.05*

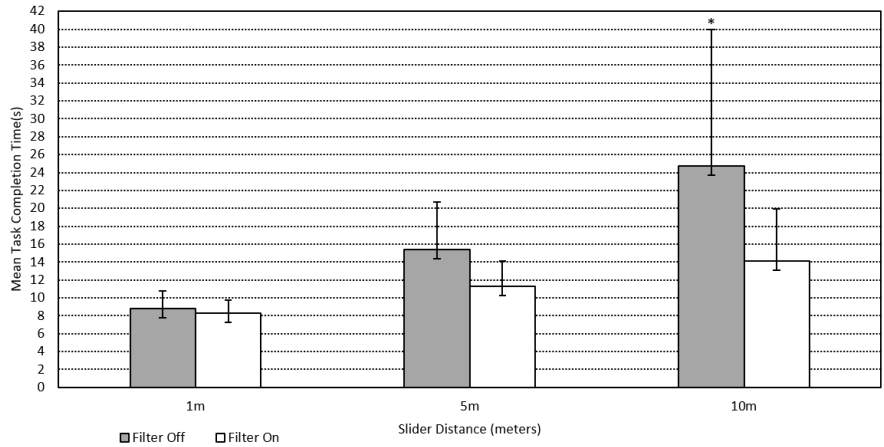

*Figure* **4** *Slider task completion mean and standard deviation*

|  | **Filter Off** | **Filter On** | **Difference** | **Reduction (%)** |
|---|---|---|---|---|
| **1m** | 8.806 | 8.254 | 0.552 | 6.27% |
| **5m** | 15.387 | 11.271 | 4.116 | 26.76% |
| **10m** | 24.700 | 14.125 | 10.573 | 42.80% |

*Table 2 Mean slider completion times with mean difference and mean % difference*

## 2.4 Button

Results for button selection-based tasks from a virtual numpad indicate that 10m ranged tasks showed highly significant improvements to task completion time with the filter enabled t(29)= 4.5569, p < 0.01. Therefore, we strongly reject the null hypothesis and find task completion time was on average 9.58 seconds faster with the filter enabled. Which was on average a 44.59% improvement in task completion time for users during 10m button tasks.

We again find significant improvement to 5m ranged tasks showing t(29)= 2.1426, p=0.0406 where we therefore reject the null hypothesis. This resulted in an average completion time of 1.72 seconds faster with the filter enabled.

However we found minimal difference in completion time for 1m ranged tasks, showing t(29)= 1.0608, p=0.2975 where we therefore accept the null hypothesis. 1m Ranged tasks resulted in an average completion time of only 0.434 seconds faster with the filter enabled.

|         | **1m**  | **5m**  | **10m**  |
|---------|---------|---------|----------|
| **p-value** | 0.2975 | 0. 0406 | 0.00008 |

*Table 3 Button task completion p-value at given ranges with a significance level of 0.05*

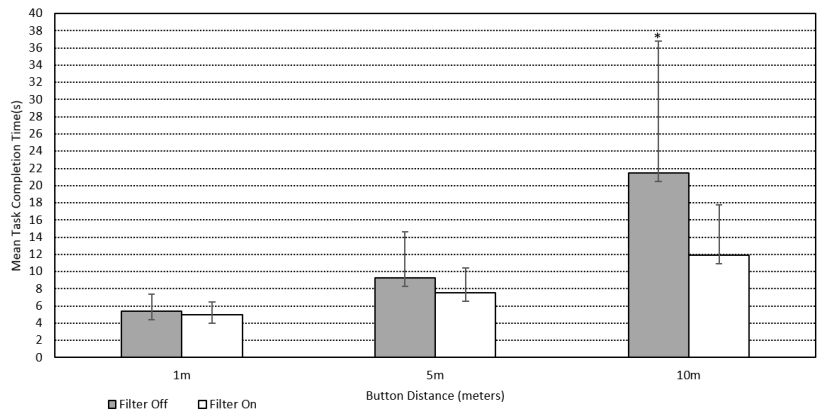

*Figure 5 Button task completion mean and standard deviation*

|        | **Filter Off** | **Filter On** | **Difference** | **Reduction (%)** |
|--------|----------------|---------------|----------------|-------------------|
| **1m**  | 5.413          | 4.978         | 0.434          | 8.02%             |
| **5m**  | 9.242          | 7.523         | 1.719          | 18.60%            |
| **10m** | 21.507         | 11.924        | 9.583          | 44.56%            |

*Table 4 Mean button completion times with mean difference and mean % difference*

## 2.5 Discussion

With the filter enabled, metrics trended towards a reduction in task completion time from 1m to 10m range tasks and on average showed a reduction in overall task completion time with the filter enabled. Showing minor fluctuations in performances at close ranges - 1m to 5m - during both slider and button input. The mean time differences between 1m and 5m with the filter enabled showed minor average task completion times, however, it is still important to discuss the benefits it had on some of the population, just because a small population doesn't exhibit statistically significant improvement in some tests doesn't mean there aren't significant advantages for a demographic.

During interactions with the filter enabled, we observe a reduction in variance among participants which is correlated with the improvement in task completion time. It is noteworthy that while there was an average improvement in task completion time, the presence of the filter ensured that task completion never exceeded the maximum observed during interactions with the filter disabled.

The correlation between the timing of the slider and button tasks was unexpected, as we see similar average task completion time improvements despite utilizing different interface elements. The slider showed average times trending very similar to the button's average completion times, in which the 10m ranged tasks 42.83% was the average improvement with the filter enabled for the slider and 44.59% improvement for the button. It should be considered however that the button interface required 4 integers to be selected over the sliders 2. This unanticipated finding raises interesting questions regarding the effectiveness of ray-based interaction for button input over a slider counterpart. It prompts the exploration into whether using a slider for text input as opposed to traditional buttons could potentially enhance the speed of text input – a known constraint in VR interfaces.

As discussed, the implementation of such a filter is low on system resources and can be enabled or disabled in real-time with a toggle at the user's discretion. Users who may not have increased times with the filter enabled could disable the filter in real-time or adjust the values to suit. Further study could be completed on the filters used with 3D GUI interfaces and how interaction could be eased with the use of ray-tracking eye focal points.